\documentclass{PoS}

\usepackage{amsmath,amssymb,mathrsfs,bm}

\def\bs#1{\boldsymbol{#1}}

\title{Solitonic Excitations In Collisions Of Superfluid Nuclei}

\ShortTitle{Solitonic Excitations In Collisions Of Superfluid Nuclei}

\author{
\speaker{Kazuyuki Sekizawa},$^a$ Piotr Magierski$^{a,b}$ and Gabriel Wlaz{\l}owski$^{a,b}$\\
\llap{$^a$}Faculty of Physics, Warsaw University of Technology\\
ulica Koszykowa 75, 00-662 Warsaw, Poland\\
\llap{$^b$}Department of Physics, University of Washington\\
Seattle, Washington 98195-1560, USA\\
E-mail: \email{sekizawa@if.pw.edu.pl}, \email{magiersk@if.pw.edu.pl}, \email{gabrielw@if.pw.edu.pl}
}


\abstract{
We investigate the role of the pairing field dynamics in low-energy heavy ion
reactions within the nuclear time-dependent density functional theory extended
to superfluid systems. Recently, we have reported on unexpectedly large effects
associated with the relative phase of the pairing field of colliding nuclei on
the reaction outcomes, such as the total kinetic energy and the fusion cross section
[P.~Magierski, K.~Sekizawa, and G.~Wlaz{\l}owski, arXiv:1611.10261 [nucl-th]].
We have elucidated that the effects are due to creation of a ``domain wall'' or
a ``solitonic structure'' of the pairing field in the neck region, which hinders energy
dissipation as well as the neck formation, leading to significant changes of the reaction
dynamics. The situation nicely mimics the one extensively studied experimentally with
ultracold atomic gases, where two clouds of superfluid atoms possessing different
phases of the pairing field are forced to merge, creating various topological excitations,
quantum vortices and solitons, as well as Josephson currents. In this paper,
we present unpublished results for a lighter system, namely, $^{44}$Ca+$^{44}$Ca.
It is shown that the pairing effects on the fusion hindrance are rather small in lighter
systems, due to a strong tendency towards fusion, which is consistent with an earlier study.
}

\FullConference{
The 26th International Nuclear Physics Conference\\
11-16 September, 2016\\
Adelaide, Australia
}

\begin{document}

%Deadline for submission: January 31st, 2017
%Less than 8 pages

\section{Introduction}

Topological excitations are among the most peculiar properties of superfluid
systems. A typical example is the existence of the quantum vortex, which was
first predicted in superfluid $^4$He by Onsager in 1949~\cite{Onsager(1949)}
and was confirmed experimentally by Vinen in 1958~\cite{Vinen(1958)}.
The quantum vortex is a manifestation of a ``winding'' of the phase of the
pairing field (the order parameter), which generates rotating supercurrents
having a normal (non-superfluid) core at the center of the vortex as a topological
defect. Nowadays it is possible to experimentally study dynamics of topological
excitations in superfluid systems with ultracold atomic gases. In the experiment
of Ref.~\cite{Ku(2016)}, for example, a cascade of solitonic excitations was
observed after a merger of two clouds possessing different phases. Namely, at first a
``domain wall'' was created, in which the superfluidity is lost as the phase
is changing rapidly, thus being a topological defect, which subsequently decays into a vortex
ring and vortex lines. The study of dynamic excitation modes of superfluid systems
is the forefront topic both experimentally and theoretically.

It is the common sense that nucleons in the majority of the atomic nuclei or
in the neutron stars are in the superfluid phase. Indeed, it has been envisaged~\cite{Anderson-Itoh(1975)}
that the pulsar glitch, a sudden spin-up of the rotational frequency, is caused
by a catastrophic ``unpinning'' of a huge number of vortices which are ``pinned''
(immobilized) by the Coulomb lattice of neutron-rich nuclei immersed in neutron
superfluid in the inner crust of neutron stars. Now, a naive question arises: does
the topological excitations of the superfluid nucleons play any role in nuclear reactions?
In the case of finite nuclei, the presence of a quantum vortex is hardly expected, since
the pairing correlations are weak, in a sense that the ratio of the pairing gap to the
Fermi energy is small, \textit{i.e.} $\Delta/\varepsilon_{\rm F} \lesssim 5\%$,
and the coherence length, typical size of a quantum vortex, becomes significantly larger
than the size of the system. Moreover, one would naively expect that the pairing in
the nucleus is so fragile that it would only affect tunneling phenomena near and below
the Coulomb barrier, like Josephson currents \cite{Dietrich(1970)}, and it would not
play important role in dissipative collisions above the barrier.

Contrary to the naive expectations, in our recent work \cite{soliton}, we have found
noticeably large effects of the pairing in low-energy heavy ion reactions. The effects
are associated with the ``phase'' of the complex pairing field, $\Delta(\bs{r})=
|\Delta(\bs{r})|e^{i\varphi(\bs{r})}$, and more precisely, with the
relative phase, $\Delta\varphi \equiv \varphi_1-\varphi_2$, between two colliding
nuclei, where $\varphi_i$ denotes the phase of the pairing field of each nucleus,
which is uniform in their ground state. The phase difference $\Delta\varphi$
triggers creation of a ``solitonic excitation'' of the pairing field in the neck region,
where the pairing is vanishing due to the phase discontinuity, which hinders energy
dissipation as well as the neck formation, leading to significant changes in the
reaction dynamics: \textit{e.g.}, total kinetic energy of the outgoing fragments in
$^{240}$Pu+$^{240}$Pu is changed up to 20~MeV and the energy necessary to fuse
two nuclei in $^{90}$Zr+$^{90}$Zr is changed by almost 30~MeV, depending on the
phase difference $\Delta\varphi$ \cite{soliton}.

It is worth noting here that although the situation nicely mimics the one studied
with ultracold atomic gases the physics of interest is quite different. In the ultracold
atomic gases, the pairing is so strong that the coherence length is on the same order
as the mean inter-particle distance, which is much smaller than the size of the system.
Due to this fact the manifestations of topological excitations, like, \textit{e.g.}, creation
and decay of a vortex ring and vortex lines, and the dynamics of Josephson currents
are better pronounced. On the other hand, in the case of nuclear reactions, the main
concern would not be dynamics of topological excitations itself, but the possible influence
on reaction mechanisms, such as dynamics of fusion, (quasi)fission, transfer reactions,
energy dissipation, collective and single-particle excitations, quantum tunneling, and so on.
Especially, the fact that the system may split after collision, due to the interplay
between nuclear forces and the Coulomb repulsion, is the unique property
of the nuclear system, which has not been studied with ultracold atomic gases.

The pairing effects in nuclear reactions have rarely been investigated to date.
The most satisfactory description is based on time-dependent
density functional theory (TDDFT) \cite{DFT1,DFT2,DFT3}. In the field of nuclear
physics, it has been developed as time-dependent mean-field theories, such as
time-dependent Hartree-Fock(-Bogoliubov) theory [TDHF(B)], with effective
two- and three-body nuclear interactions (for recent reviews, see
Refs.~\cite{Simenel(review),Nakatsukasa(review)}). To perform a full TDHFB
calculation is still computationally challenging. Thus, the possible pairing effects
on the reaction dynamics were investigated with simplified approaches
\cite{Blocki-Fkocard(1976),Ebata(2010)}. Very recently, the first attempt has been
reported in Ref.~\cite{HS(TDHFB)}, where the effects of the phase difference
in head-on collisions of $^{20}$O+$^{20}$O were investigated based on TDHFB.
From the results, a vestige of the repulsive effect of the phase difference was
indeed seen in collision trajectories, although the magnitude is very small,
as compared to our results for Zr+Zr and Pu+Pu systems. In order to clarify
this issue, in this article we examine
the effects in collisions of relatively light nuclei, $^{44}$Ca+$^{44}$Ca,
at energies around the Coulomb barrier.

This article is organized as follows.
In Sec.~\ref{sec:theory}, we briefly summarize our theoretical framework.
In Sec.~\ref{sec:results}, we show results of TDSLDA calculations for the $^{44}$Ca+$^{44}$Ca reaction.
In Sec.~\ref{sec:summary}, a short summary is given.

\section{Theoretical Framework}{\label{sec:theory}}

We use a microscopic framework based on TDDFT, which is capable of
describing reaction dynamics, taking explicitly into account nucleonic degrees of freedom. We utilize
a local treatment of superfluid TDDFT known as time-dependent superfluid
local density approximation (TDSLDA). The feasibility of the approach has been
tested for describing the dynamics of strongly correlated Fermionic systems in
both ultracold atomic gases \cite{Bulgac(2009),Bulgac(2011),UFG(2012),
Bulgac(2012),Bulgac(2013),Bulgac(2014),Gabriel(2015)} and in nuclear systems
\cite{Stetcu(2011),Stetcu(2015),Pu-fission,Piotr(2016)}. To simulate heavy
ion reactions, we have extended a computational code that we used to study
dynamics of a quantum vortex in the presence of a nuclear impurity in the
inner crust of neutron stars \cite{vortex}. Here we briefly summarize the
theoretical and computational aspects of the framework. (We refer readers
to Refs.~\cite{soliton,vortex}, for more details.)

We numerically solve the TDSLDA equations
%(formally equivalent to TDHFB or time-dependent Bogoliubov-de Gennes equations),
\begin{equation}
i\hbar\frac{\partial}{\partial t}
\begin{pmatrix}
u_i(\bs{r})\\
v_i(\bs{r})
\end{pmatrix}
=
\begin{pmatrix}
h(\bs{r})              & \Delta(\bs{r}) \\
\Delta^*(\bs{r}) & -h(\bs{r})
\end{pmatrix}
\begin{pmatrix}
u_i(\bs{r})\\
v_i(\bs{r})
\end{pmatrix},
\end{equation}
where $u_i(\bs{r})$ and $v_i(\bs{r})$ are the quasiparticle wave functions.
$h(\bs{r})$ is the single-particle Hamiltonian and $\Delta(\bs{r})$ is the pairing
field, which are derived from appropriate functional derivatives of an energy density
functional (EDF), $\mathcal{E}(\bs{r})=\mathcal{E}_0(\bs{r})+\mathcal{E}_{\rm pair}(\bs{r})$.
For the normal part, $\mathcal{E}_0(\bs{r})$, we use FaNDF$^0$ functional proposed
by Fayans \cite{Fayans}. In the present study we neglect the spin-orbit term in the
functional, which allows to construct a highly efficient TDSLDA solver \cite{vortex}
which works on hundreds of GPUs with almost perfect scalability. We supplement
the Fayans EDF with a local pairing functional,
\begin{equation}
\mathcal{E}_{\rm pair}(\bs{r})=-g\bigl[ |\nu_n(\bs{r})|^2 + |\nu_p(\bs{r})|^2 \bigr],
\end{equation}
where $\nu_{n(p)}(\bs{r})$ are neutron (proton) anomalous densities.
The local treatment of the pairing requires a regularization, since the
anomalous density is divergent, $\nu(\bs{r},\bs{r}') \propto
\frac{1}{|\bs{r}-\bs{r}'|}\rightarrow\infty$ for the limit $|\bs{r}-\bs{r}'|
\rightarrow 0$ \cite{Bulgac(2001)}. We apply a regularization procedure
of Ref.~\cite{Bulgac(2002)2}. Densities of neutrons ($q=n$) and protons
($q=p$) are then evaluated as
$\rho_q(\bs{r})=2\,\tilde{\sum}_{i\in q}|v_i(\bs{r})|^2$,
$\tau_q(\bs{r})=2\,\tilde{\sum}_{i\in q}|\bs{\nabla}v_i(\bs{r})|^2$,
$\nu_q(\bs{r})=\tilde{\sum}_{i\in q} v_i^*(\bs{r})u_i(\bs{r})$ 
(the factor of two stands for the spin degree of freedom), where $\tilde{\sum}_{i\in q}$
takes summation over positive quasiparticle energy states defined at $t=0$ smaller
than a cutoff, $0 \le E_i \le E_c$. We note that the pairing field $\Delta_q(\bs{r})$
does not depend on the regularization procedure.

The quasiparticle wave functions are represented on three-dimensional
Cartesian spatial lattice (without symmetry restrictions) with periodic boundary
conditions. A box of $80\;{\rm fm}\times25\;{\rm fm}\times25\;{\rm fm}$
with lattice spacing of 1.25~fm was used to simulate head-on collisions.
The spatial derivatives and the Coulomb potential are computed employing
Fourier transforms. The initial wave function of projectile and
target nuclei placed with a certain distance within the computational box is
prepared using Shifted Conjugate Orthogonal Conjugate Gradient (COCG)
method \cite{COCG}, combined with a direct diagonalization of the Hamiltonian
matrix. We used an external potential to keep two nuclei at rest during the
self-consistent iterations (to compensate the Coulomb repulsion). After getting
a convergent solution, the phase difference was dynamically imprinted with
a constant external potential for half of the box. We also used an external
potential to boost two nuclei. For time evolution, the Trotter-Suzuki decomposition
was used with a single predictor-corrector step. The time step was set to
$\Delta t \simeq 0.038$~fm/$c$. The cutoff energy for the pairing regularization
was set to $E_c=100$~MeV. The corresponding numbers of quasiparticle wave
functions for neutrons and protons within the initial cutoff energy were 8,740 and 7,723,
respectively. These settings ensured the stable time evolution within the 
intervals exceeding 12,000~fm/$c$.

We have neglected the spin-orbit interaction which, although
crucial for a proper description of nuclear structure (\textit{i.e.}, shell structure and
deformation) and energy dissipation in low-energy heavy ion reactions \cite{Umar(1986)},
does not influence the mechanism of described effect.
It is worth emphasizing here that the impact of the pairing phase difference on
the reaction dynamics may be captured even without the spin-orbit interaction.
The energy cost to build the ``domain wall'' is given by (derived from
Ginzburg-Landau theory) \cite{soliton}
\begin{equation}
E_j = \frac{S}{L}\frac{\hbar^2}{2m}n_s\sin^2\frac{\Delta\varphi}{2},
\label{Eq:E_junction}
\end{equation}
where $S$ is the surface area of the wall, $L$ is the length over which the phase
varies, $m$ is the nucleon mass, and $n_s$ is the superfluid density. The main
ingredients to have reasonable values of $S,L,n_s$, \textit{i.e.} the radius of
the nucleus, the Fermi energy and the pairing strength, are correctly described
even without the spin-orbit interaction. Thus, nevertheless the interaction is not
fully realistic, our framework is enough to study the possible impact of the pairing
field dynamics in low-energy heavy ion reactions. Moreover, apart from the inevitable
increase of enormous computation costs, inclusion of the spin-orbit coupling will
also increase the complexity of the reaction mechanism (see, \textit{e.g.},
Refs.~\cite{Maruhn(2006),Iwata(2011)}). Therefore, to clearly underpin the
possible effects of the pairing, to neglect the spin-orbit coupling would be
rational, as a first step.

\section{Pairing Effects in a Lighter System---$^{44}$Ca+$^{44}$Ca case}{\label{sec:results}}

%&&&&&&&&&&&&&&&&&&&&&&&&&&&&&&&&&&&&&&&&&&&&&&&&&&
\begin{figure}[t]
\begin{center}
\includegraphics[width=\textwidth]{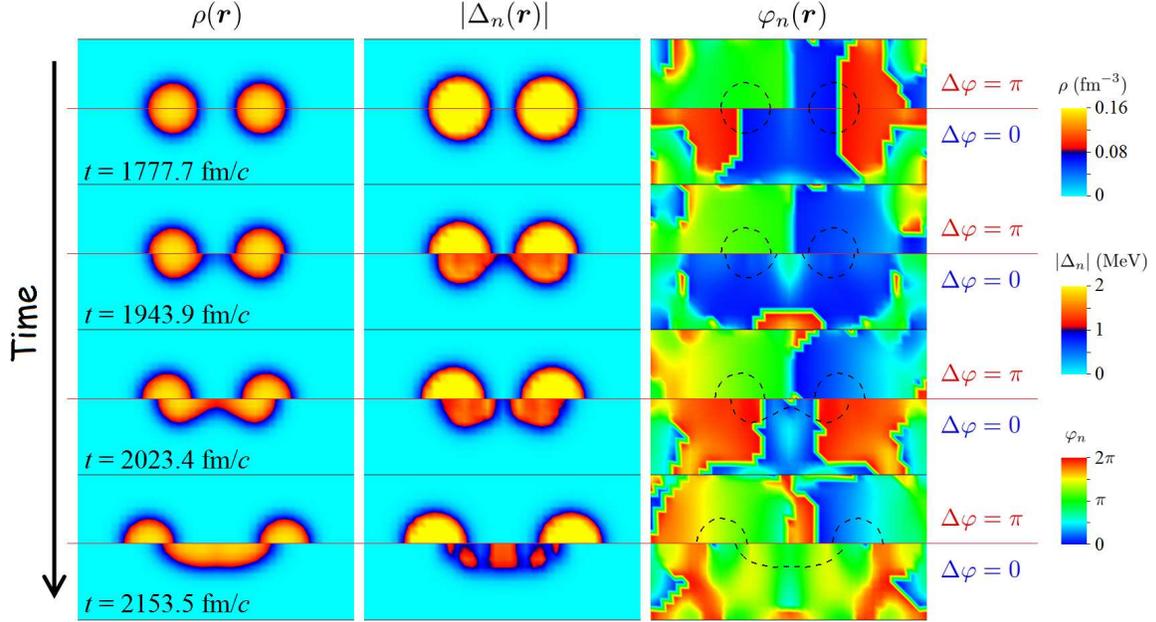}
\end{center}\vspace{-3mm}
\caption{
Results of the TDSLDA calculations for the $^{44}$Ca+$^{44}$Ca reaction
at $E \simeq 1.09V_{\rm Bass}$. $V_{\rm Bass}$ ($\simeq 48$~MeV) is
the phenomenological fusion barrier \cite{Bass(1974)}. In the left, middle
and right columns, respectively, the total density $\rho(\bs{r})$, the absolute
value and the phase of the neutron pairing field, $\Delta_n(\bs{r})=|\Delta_n(\bs{r})|
e^{i\varphi_n(\bs{r})}$, are shown, as a cross section on the reaction plane.
In the upper (lower) half of each panel, the $\Delta\varphi=\pi$ (0) case
is presented. Each row corresponds to a different time. In the right column,
a contour of density, $\rho_0/2=0.08$~fm$^{-3}$, is depicted by dashed
lines. Note that blue and red colors in the right column, corresponding to
$\varphi_n(\bs{r})=0$ and $2\pi$, respectively, are equivalent.
}
\label{FIG:Ca+Ca}
\end{figure}
%&&&&&&&&&&&&&&&&&&&&&&&&&&&&&&&&&&&&&&&&&&&&&&&&&&

In order to investigate the pairing effects in a lighter system, we selected
$^{44}$Ca+$^{44}$Ca system. Since the proton number $Z=20$ is a magic
number even without the spin-orbit interaction, only neutrons are in superfluid
phase. Here, let us first analyze the main effects of the pairing phase difference
on the reaction dynamics. In Fig.~\ref{FIG:Ca+Ca}, an illustrative example
is shown for two cases, $\Delta\varphi=0$ and $\pi$. At this collision energy
($E=1.09V_{\rm Bass}$), the $\Delta\varphi=0$ case resulted in fusion,
whereas in the $\Delta\varphi=\pi$ case binary fragments were observed.

The observed difference of the reaction dynamics is essentially caused by dynamic
effects. The crucial difference can be seen in the second row of the figure
($t=1943.9$~fm/$c$). The density distribution $\rho(\bs{r})$ (left) exhibits a 
subtle neck between two nuclei in the $\Delta\varphi=0$ case (lower part).
This fact can be understood as follows. The ``precursor'' of the neck is expected to be mainly
formed by the neutrons near the Fermi level, which also play a predominant role in
the pairing phenomena. On the other hand, any spatial change of the phase of the
pairing field induces a supercurrent, as its velocity is proportional to the gradient of
the phase, \textit{i.e.}, $\bs{v}_s(\bs{r}) = (\hbar/2m)\bs{\nabla}\varphi(\bs{r})$.
If the phase exhibits large variations in space, the supercurrent would be very large,
which is clearly unfavorable. The system thus chooses to become normal in
such a region, where the phase is changing steeply. This is the reason why we
have observed a ``domain wall'', where the pairing field is vanishing in the neck
region, especially in the $\Delta\varphi=\pi$ case \cite{soliton} (see also
Fig.~\ref{FIG:Ca+Ca}). Therefore, the phase difference prevents the superfluid
neutrons to take part in the formation of the precursor of the neck, which resulted
in the dramatic change of the reaction dynamics, as shown in Fig.~\ref{FIG:Ca+Ca}.

%&&&&&&&&&&&&&&&&&&&&&&&&&&&&&&&&&&&&&&&&&&&&&&&&&&
\begin{figure}[t]
\begin{center}
\includegraphics[width=\textwidth]{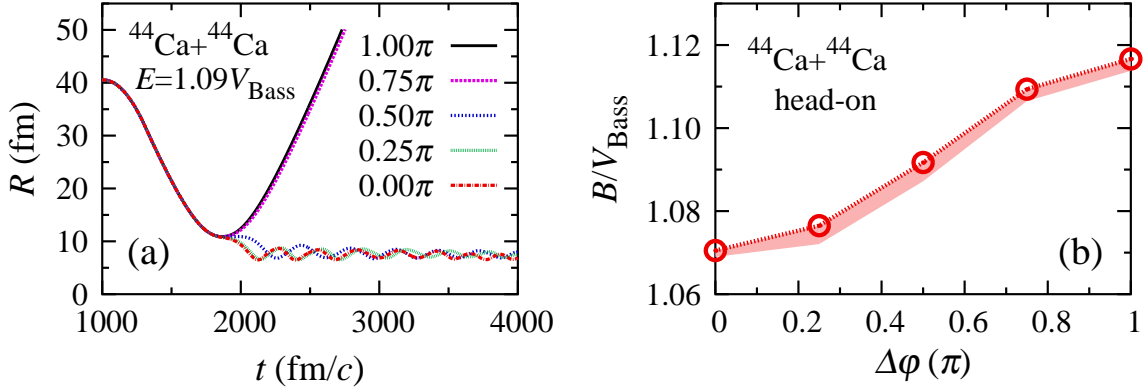}
\end{center}\vspace{-3mm}
\caption{
Results of the TDSLDA calculations for the $^{44}$Ca+$^{44}$Ca reaction.
(a) Relative distance $R(t)$ in the $^{44}$Ca+$^{44}$Ca reaction at
$E \simeq 1.09V_{\rm Bass}$ for various relative phase $\Delta\varphi$
is shown as a function of time. $V_{\rm Bass}$ ($\simeq 48$~MeV) is the
phenomenological fusion barrier \cite{Bass(1974)}. (b) The fusion threshold
energy $B/V_{\rm Bass}$ is shown as a function of the relative phase $\Delta\varphi$.
}
\label{FIG:Fusion}
\end{figure}
%&&&&&&&&&&&&&&&&&&&&&&&&&&&&&&&&&&&&&&&&&&&&&&&&&&

In Fig.~\ref{FIG:Fusion}~(a), we show the relative distance $R(t)$
between the two colliding nuclei in the $^{44}$Ca+$^{44}$Ca reaction
at $E=1.09V_{\rm Bass}$ as a function of time for various phase differences.
In this case, we observe fusion for $\Delta\varphi \le \pi/2$. The fusion reaction
does not occur for $\Delta\varphi=3\pi/4$ and $\pi$ due to the hindered neck
formation. In this way, the system requires additional energy to fuse two colliding
nuclei, which increases with the phase difference. By repeating the simulations with
different collision energies, we searched for the minimum energy at which the fusion reaction
takes place. In Fig.~\ref{FIG:Fusion}~(b), the obtained fusion threshold energy
$B$ is shown as a function of the phase difference $\Delta\varphi$. The filled
area indicates the uncertainty due to finite collision energy steps ($\lesssim 220$~keV).
From the figure, we find a change of the fusion threshold energy, up to about 5\%
of the barrier (2.3~MeV). Taking an average over the phase difference, we obtain
an effective barrier increase of $E_{\rm extra} = \frac{1}{\pi}\int_0^\pi
[B(\Delta\varphi)-B(0)] d(\Delta\varphi) \approx 1.3$~MeV. Interestingly,
we find a ($\sin^2\frac{\Delta\varphi}{2}$)-like pattern in Fig.~\ref{FIG:Fusion}~(b),
the same dependence as the energy of the domain wall [cf.~Eq.~(\ref{Eq:E_junction})],
which was not present in a heavier system, Zr+Zr \cite{soliton}. It is worth emphasizing
here that physics of the observed effect cannot be explained as the nuclear Josephson effect, since
the Josephson current is proportional to $\sin\Delta\varphi$, which clearly
fails to explain observed $\sin^2\frac{\Delta\varphi}{2}$ pattern \cite{soliton}.

\section{Summary}{\label{sec:summary}}

We have performed three-dimensional, microscopic, dynamic simulations
of low-energy heavy ion reactions based on time-dependent density functional
theory extended to superfluid systems. We have investigated the effects of the
relative phase of the complex pairing field of colliding nuclei on the reaction
dynamics in a relatively light system, $^{44}$Ca+$^{44}$Ca. We have found
that the fusion reaction is hindered by the phase difference, due to the suppressed
neck formation, as was observed in a heavier system, Zr+Zr \cite{soliton}.
However, the magnitude of the effective barrier increase does not exceed several percent
of the Coulomb barrier, consistent with the earlier study \cite{HS(TDHFB)}.
In order to make a quantitative prediction, realistic simulations including
the spin-orbit coupling are mandatory.

\begin{acknowledgments}
This work was supported by the Polish National Science Center (NCN)
under Contracts No. UMO-2013/08/A/ST3/00708. The code used for
generation of initial states was developed under grant of Polish NCN
under Contracts No. UMO-2014/13/D/ST3/01940. Calculations have
been performed at HA-PACS (PACS-VIII) system---resources provided
by Interdisciplinary Computational Science Program in Center for
Computational Sciences, University of Tsukuba.
\end{acknowledgments}

\end{document}